# Rapid Generation of a Macroscopic Schrödinger Cat State of Atoms with Parity-Independent Orientation

Jinyang Li[1], Gregório R. M. da Silva[1], Schuyler Kain[1], Selim M. Shahriar[1,2]

[1] Department of Physics and Astronomy, Northwestern University, Evanston, IL 60208, USA

[2] Department of ECE, Northwestern University, Evanston, IL 60208, USA

## Abstract

We show that using the process of one-axis-twist squeezing in an echo configuration, it is possible to control the orientation of the macroscopic magnetic moment of a large number of atoms by manipulating the quantum state of a single atom that is physically isolated from the ensemble. With this control technique, it is also possible to entangle an ensemble with a single atom deterministically, which mimics the thought experiment known as the Schrödinger's cat. In addition, this technique would make it possible to generate a mesoscopic Schrödinger cat state for a large number of atoms far more rapidly that the conventional process for generating such a state, with an orientation that is independent of the parity of the number of atoms.  Apart from the echo configuration, we have also investigated the behavior of one-axis-twist squeezing for some special values of the squeezing parameter. We find that the squeezing propagator can be expressed as the sum of $n$ rotation operators if the product of $n$ and the squeezing parameter equals pi, where $n$ is a non-zero integer. A direct consequence of this property of one-axis-twist squeezing is that there is a hidden order in a squeezed state generated under this condition even if its Husimi quasi-probability distribution looks irregular.

# 1. Introduction

Reconciliation of quantum mechanics (QM) or, more generally, the standard model, with general relativity has not yet been achieved. Aside from this challenge, QM is viewed as the undisputed theory of how nature works. The law of QM applies on all scales. Just as the center of mass of an electron or an atom is expected to obey the Schrödinger equation (in the non-relativistic limit), so should the center of mass of a large object such as a glass marble with a cm-scale radius. Modern experiments routinely produce quantum superposition of different states of microscopic objects. For example, a recent experiment has produced a quantum superposition of a Rb atom separated by 54 cm [1]. Given that an atom is very small at the scale of human perception, such a superposition does not seem to elicit a sense of wonder at how QM differs from common sense understanding of the laws of nature. On the other hand, if it were possible to create such a superposition of a cm-scale marble, it would fundamentally alter our perspective on nature. Such a superposition of a macroscopic object is so radical that in his famous paper [2] Schrödinger, after describing a plausible scenario for creating a superposition of a cat being alive and dead, implied that quantum theory is not complete and fully understood. In fact, there are many current theories [3, 4, 5, 6, 7] that imply that QM, as formulated, is incomplete, and additional modification thereof is warranted when applying to macroscopic objects. It is generally understood that a spatially separated superposition of a macroscopic object would decay to one of the states extremely rapidly. From the point of view of microscopic quantum theory, such a collapse is expected to result from the decoherence of internal constituent particles that are at a finite temperature. However, other theories, such as the one posed independently by Diósi and Penrose [3, 4, 5, 6, 7] imply that such a collapse may occur even if the decoherence of the internal particles is not taken into account. In order to address the question of whether QM indeed allows the creation of a spatially separated superposition of a macroscopic object, one must study such a process experimentally. One way to create spatially separated superpositions of macroscopic objects is to produce a quantum superposition of a cavity mirror on a cantilever with the idea of optomechanics [8, 9, 10]. However, the two quantum states can be only separated by a small distance with this method, and thermal noise is a big challenge for such a method. Recently, we had proposed a scheme [11, 12] that can, in principle, produce a spatially separated superposition of as many as 140 billion atoms (limited by the maximum number of atoms caught in a magneto-optic trap to date [13]). However, this process is probabilistic since the nature of the superposition depends on the parity of the number of atoms involved. Besides, a relatively long atom-cavity interaction time is required for this scheme, significantly lowering the success rate, especially for a large number of atoms. In this paper, we propose a method using one-axis-twist squeezing in an echo configuration to produce a Schrödinger cat state in a deterministic manner and with a much shorter atom-cavity interaction time.

It should also be noted that the scheme proposed by Schrödinger for creating a superposition of a dead cat and a live cat is highly misleading in this context, because the scheme involves the detection of a quantum state, inducing a collapse of the superposition. To illustrate more explicitly, let us consider a modern and simpler version of the experiment that would mimic the idea presented by Schrödinger. Imagine the creation of a photonic quantum bit that is in a linear superposition of two polarization states. We then send this photon through a polarizing beam splitter, and place a single photon detector in one port of the beam splitter. If the detector sees a photon, it produces a signal that moves a cat sitting on a cart on a rail to a different location, B. If it does not, then the cat remains in the original position A. The implication in Schrödinger's paper (modified for the scenario presented here) is that this creates a superposition of the cat in two different positions, A and B, at the same time. However, this is not at all correct. The detection of the photon constitutes a measurement process, which collapses the quantum state of the photon. As such, the process does not lead to a superposition of the cat in two different places. All we can see from this thought experiment is that if we repeat the process many times, we will, with equal probability, find the cat either in position A or position B. This is hardly radical, since a classical probabilistic model would lead to the same conclusion. Thus, in order to create the scenario that Schrödinger viewed to be problematic, implying incompleteness of quantum mechanics, what is needed is a situation that would produce a superposition of the cat being in position A or B, without any intervening measurement process. In this paper, we describe a technique that can indeed a achieve this goal in a deterministic manner, for an ensemble of a large number of atoms, by manipulating the quantum state of a single atom that is not physically close to this ensemble. Experimental realization of such a scheme would provide a true test of whether it is indeed possible to create a spatially separated superposition of a macroscopic object by entangling it to a microscopic particle, or what the limit of such a process might be.

In addition to the echo configuration, we have also investigated the behavior of one-axis-twist squeezing for some special values of squeezing parameter $\mu$. We find that the squeezing propagator can be expressed as the sum of $n$ rotation operators if $\mu = \pi/n$, where $n$ is a non-zero integer. A direct consequence of this property of one-axis-twist squeezing is that there is a hidden order in a squeezed state when $\mu = \pi/n$ even if its Husimi quasi-probability distribution looks unregular.

The rest of this paper is organized as follows. In Sec. 2, we demonstrate the technique to control the orientation of a coherent spins state with one or a few atoms. In Sec. 3, we show the application of this technique to the generation of Schrödinger cat states. In Sec. 4, we prove that a OATS propagator can be expressed as a sum of $n$ rotation operators if $\mu = \pi/n$.

## 2. Control of the orientation of a coherent spin state with one or a few atoms

To facilitate the exposition of the schemes presented here, it is convenient to summarize first the relevant notations. A two-level atom can be modeled as a spin-1/2 spinor, with the spin operator denoted as $\mathbf{s} = (s_x, s_y, s_z)$, and the two eigenstates of $s_z$ denoted as $\{|\uparrow\rangle, |\downarrow\rangle\}$, with the eigenvalues of $\{1/2, -1/2\}$. The spin operator for an ensemble of atoms can be expressed as $\mathbf{S} = \sum_{j=1}^{N} \mathbf{s}_j$, where $N$ is the number of atoms and $\mathbf{s}_j$ is the spin operator of the $j$-th atom. The state of this atom can be described by a point on the Bloch sphere. A point on the Bloch sphere can be characterized by the polar angle $\theta$ and the azimuthal angle $\phi$. The $j^{\text{th}}$ atom in the state corresponding to such a point on the Bloch sphere is defined as $|\theta, \phi\rangle_j \equiv \cos(\theta/2)|\uparrow\rangle_j + e^{i\phi} \sin(\theta/2)|\downarrow\rangle_j$. A coherent spin state (CSS) [14, 15] characterized by the parameters $\theta$ and $\phi$ is defined as a state of $N$ atoms with each atom in the state $|\theta, \phi\rangle_j$, that is $|\theta, \phi\rangle \equiv \bigotimes_{j=1}^{N} |\theta, \phi\rangle_j$.

The primary technology that enables this process is one-axis-twist squeezing (OATS) [16, 17, 18, 19, 20]. It has been shown previously that OATS can enhance the sensitivity of quantum sensors by either suppressing the quantum noise [16, 21], or magnifying the quantum phase shift [11, 22, 23, 24]. It has been found that that OATS with the squeezing parameter $\mu = \pi/2$ can produce a so-called Schrödinger cat state, which is the maximally entangled state consisting of two coherent spin states (CSSs) in opposite directions (e.g., $+z$ direction and $-z$ direction) [25, 26, 27, 28]. The Schrödinger cat state plays an important role in quantum metrology because in principle a quantum sensor employing the Schrödinger cat state can reach the Heisenberg limit [12, 29].

The concepts presented in this paper are closely related to the protocols we had proposed previously for creating the Schrödinger cat state for metrological applications. Specifically, we show that OATS can be used for controlling the orientation of a CSS with only one atom, to be denoted as the control atom, that is physically separated from the ensemble. We start by considering the more general case where there is an arbitrary number of control atoms. We denote the control atoms as the control group and the atoms in the CSS as the target group. The $z$-component of the spin operator is denoted as $\tilde{S}_z$ for the control group and $S_z$ for the target group. The protocol using OATS for generating entanglement between these two groups of atoms simply consists of two steps. First, all the atoms undergo the OATS process, and then only the target group undergoes the inversion of the OATS process.

In practice, the atoms used can be alkali atoms which are modeled as two-level systems, and the OATS process can be realized with an optical cavity [17, 18, 19]. The experimental scheme of this protocol is illustrated schematically in Figure 1. The atoms in the control group can be either confined in one dipole trap or confined in individual dipole traps (the latter is the case shown in Figure 1), depending on the purpose. The target group is confined in another dipole trap. It is assumed that the density and the temperature of the atoms in either group are low enough so that interactions among the atoms can be ignored. We also assume that the phase space density remains above the threshold for Bose condensation. During the OATS process, all atoms are exposed to the squeezing light in the cavity. Prior to the inversion of the OATS process, the control group is removed adiabatically from the cavity by moving the dipole trap, while the target group remains in the cavity.

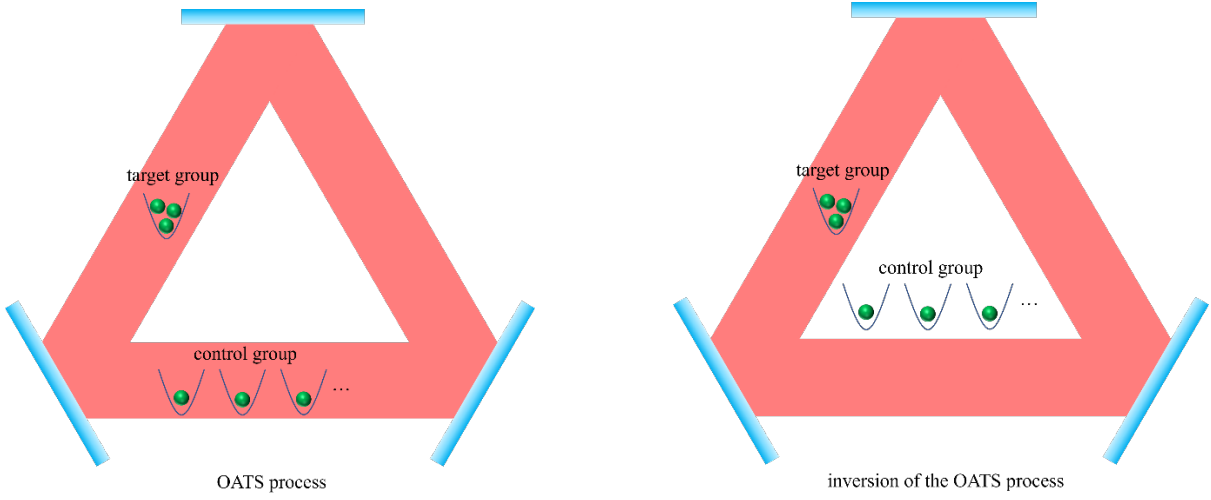

Figure 1 Schematic illustration of the process for manipulating the target group with the control group. The target group is confined in a dipole trap and each atom in the control group is confined in an individual dipole trap. During the OATS process, both the target group and the control group are exposed to the squeezing light, while the control group is moved out of cavity prior to the inverse of the OATS process,

Under ideal conditions assumed here, the Hamiltonian for the OATS process for the control (target) group can be expressed as $H = \chi \tilde{S}_z^2$ ($H = \chi S_z^2$), where the parameter $\chi$ represents the strength of the squeezing process. The sign of the Hamiltonian is inverted during the unsqueezing process; such inversion can be realized by changing the sign of the detuning of the probe beam input to the cavity with respect to the cavity resonant frequency. Thus, if we assume that the duration, $\tau$, is the same for the squeezing and unsqueezing processes, and define the squeezing parameter as $\mu \equiv \chi \tau$, then the propagator describing this two-step process can be expressed as $\exp(i\mu S_z^2)\exp(-i\mu(S_z + \tilde{S}_z)^2)$, which equals $\exp(-i2\mu S_z \tilde{S}_z)\exp(-i\mu \tilde{S}_z^2)$, since $S_z$ and $\tilde{S}_z$ commute. Assume that the target group is initially in a CSS $|\theta, \phi\rangle$ and the control group is in an arbitrary superposition of eigenstates of $\tilde{S}_z$, expressed as $\sum_m c_m |m\rangle$. After the system undergoes the protocol, the final state is calculated to be

$$e^{-i2\mu S_z \tilde{S}_z} e^{-i\mu \tilde{S}_z^2} \sum_m c_m |m\rangle \otimes |\theta,\phi\rangle = \sum_m c_m |m\rangle \otimes e^{-i2\mu m S_z} e^{-i\mu m^2} |\theta,\phi\rangle$$
$$= \sum_m c_m e^{-i\mu m(N+m)} |m\rangle \otimes |\theta,\phi+2m\mu\rangle \tag{1}$$

where $N$ is the number of atoms in the target group. We can see that if the control group is in the eigenstate $|m\rangle$, the final state of the target group is $|\theta,\phi+2m\mu\rangle$, whose orientation is determined by the value of $m$. It should be noted that this result is independent of the parity of the number of atoms in the target group. We can also see that if the control group is in a superposition of eigenstates of $S_z$, the final state involves entanglement between the control group and the target group. Figure 2 shows an example of the final state in terms of the Husimi quasi-probability distribution (QPD) [16] in the case where the control group starts in a superposition of $|m_1\rangle$ and $|m_2\rangle$ and the target group starts as CSS $|\theta,0\rangle$. The final state is in a superposition of $|m_1\rangle \otimes |\theta,2m_1\mu\rangle$ and $|m_2\rangle \otimes |\theta,2m_2\mu\rangle$. Specifically, we have considered here a case where the control group contains only three atoms, with $|m_1\rangle = |3/2\rangle$ (i.e., $|\uparrow\uparrow\uparrow\rangle$) and $|m_2\rangle = |1/2\rangle$ (i.e., $|\uparrow\uparrow\downarrow\rangle$). Note that each of these two eigenstates of the control group can be produced from the other via inversion of the spin of only one atom. The target group starts as CSS $|\pi/2,0\rangle$, which corresponds simply to the state where all atoms are aligned in the x-direction. Such a state can be produced, for example, via the application of a $\pi/2$ pulse around the y-axis to an ensemble of

atoms where each is in state $|\uparrow\rangle$. The final state is a superposition of $|\uparrow\uparrow\uparrow\rangle \otimes |\pi/2, 3\pi/5\rangle$ and $|\uparrow\uparrow\downarrow\rangle \otimes |\pi/2, \pi/5\rangle$.

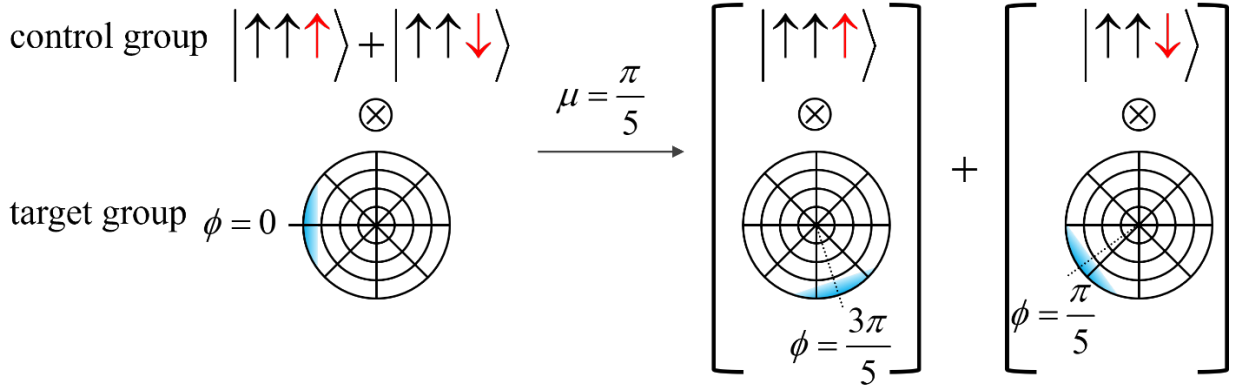

Figure 2 Example of the final state of the protocol demonstrated with the Husimi quasi-probability distribution in the case where the control group starts as a superposition of |↑↑↑⟩ and |↑↑↓⟩ (i.e., | 3/2⟩ and | 1/2⟩ in terms of the eigenvalue of $\tilde{S}_z$) and the target group starts in CSS |π/2,0⟩. The final state is a superposition of |↑↑↑⟩ ⊗ |π/2, 3π/5⟩ and |↑↑↓⟩ ⊗ |π/2, π/5⟩.

Of course, it is obvious that if we use only one control atom, in an equal superposition of state $|\uparrow\rangle$ (i.e., $|1/2\rangle$) and state $|\downarrow\rangle$ (i.e., $|-1/2\rangle$), and initially prepare the target group in CSS $|\theta, 0\rangle$, then the final state would be an equal superposition of $|\uparrow\rangle \otimes |\theta, \mu\rangle$ and $|\downarrow\rangle \otimes |\theta, -\mu\rangle$. Thus, the quantum state of the single atom can be used to control the orientation of the net spin alignment of a macroscopic ensemble.

## 3. Deterministic way to produce a Schrödinger cat state

In the field of atom interferometry, it is common to associate state $|\uparrow\rangle$ and state $|\downarrow\rangle$ of an atom with different linear momenta. We have shown [11, 12] that it is possible to produce a Schrödinger cat state $(|\Uparrow\rangle + |\Downarrow\rangle)/\sqrt{2}$, where $|\Uparrow(\Downarrow)\rangle$ denotes the CSS where all the atoms are in state $|\uparrow(\downarrow)\rangle$, with CSS $|\Uparrow\rangle$ and CSS $|\Downarrow\rangle$ associated with different linear momenta. Therefore, after a sufficiently long time, the two CSSs can be spatially separated. However, whether this method can produce this Schrödinger cat state depends on the parity of the atom number, which cannot be known for billions of atoms with the present technology. With the protocol described above, we can produce such a Schrödinger cat state deterministically by entangling the target group to a single atom in a superposition of state $|\uparrow\rangle$ and state $|\downarrow\rangle$, which is akin to the scheme proposed by Schrödinger. The process can be described as follows. First, a $\pi/2$ rotation around the $y$ axis is applied to the target group initially in the CSS $|\Uparrow\rangle$. At the end of this step, the target group will be in CSS $|\pi/2,0\rangle$. Then the target group together with the control group in state $(|\uparrow\rangle+|\downarrow\rangle)/\sqrt{2}$ undergoes the two-step protocol described above, with $\mu=\pi/2$. The resulting state will be an equal superposition of $|\uparrow\rangle\otimes|\pi/2,\pi/2\rangle$ and $|\downarrow\rangle\otimes|\pi/2,-\pi/2\rangle$. Next, another $\pi/2$ pulse around the $x$ axis is applied to the target group so that the final state is an equal superposition of $|\uparrow\rangle\otimes|\Uparrow\rangle$ and $|\downarrow\rangle\otimes|\Downarrow\rangle$. After a sufficiently long time, the two CSSs $|\Uparrow\rangle$ and $|\Downarrow\rangle$ will be spatially separated. It should be noted that this result does not depend on the parity of the atom number, and is thus deterministic. It has been shown [13] that as many as 140 billion atoms can be trapped with a magneto optical trap, with a cloud diameter larger than one cm. As such, the target ensemble can be as large as this, thus realizing what Schrödinger viewed as evidence of incompleteness of quantum mechanism.

We note that the state $(|\uparrow\rangle\otimes|\Uparrow\rangle+|\downarrow\rangle\otimes|\Downarrow\rangle)/\sqrt{2}$ can be viewed as the standard Schrödinger cat state $(|\Uparrow\rangle+|\Downarrow\rangle)/\sqrt{2}$ if the control atom is viewed as being part of the target group.

It has been proposed to use OATS with $\mu = \pi/2$ to generate the Schrödinger cat state for quantum sensing [12, 29]. However, the orientation of the Schrödinger cat state generated with the approach described in Refs. [12] and [29] depends on the parity of $N$, making the maximum achievable sensitivity the Heisenberg limit within a factor of $\sqrt{2}$ if the value of $N$ is unknown [11]. With the technique described above, we can control the orientation of the Schrödinger cat state without knowing the parity of $N$, and thus make the maximum achievable sensitivity the Heisenberg limit.

We next consider a more general case where the control group consists of a small number, $N'$, of atoms. These atoms are placed in the Schrödinger cat state $\left(|\Uparrow\rangle_c + |\Downarrow\rangle_c\right)/\sqrt{2}$, where the subscript 'c' indicates that this state is for the control group. This can be achieved deterministically using, for example, Rydberg-blockade induced CNOT gates among neighboring atoms [30, 31]. This step is relatively reliable because only a few atoms are involved. Next, the two-step protocol described in Eq. (1) is applied to the target group in CSS $|\theta,\phi\rangle$ to create the entangled state $\left(|\Uparrow\rangle_c \otimes |\theta,\phi+\mu N'\rangle_t + |\Downarrow\rangle_c \otimes |\theta,\phi-\mu N'\rangle_t\right)/\sqrt{2}$, where the subscript 't' refers to the target atoms. For $\mu N' = \pi/2$, this technique, when applied to the atomic beam splitter describe above, can be used generate the state $\left(|\Uparrow\rangle_c \otimes |\Uparrow\rangle_t + |\Downarrow\rangle_c \otimes |\Downarrow\rangle_t\right)/\sqrt{2}$, where the target group is in a superposition of two physically separated macroscopic states. When $N'=1$, this case is identical to the one described earlier, requiring $\mu = \pi/2$. However, this generalized result shows that it is possible to create such a state for much smaller values of $\mu$, namely $\pi/2N'$. In the experimental context, this is important, since achieving the conditions corresponding to $\mu = \pi/2$ for a large number of atoms can be difficult, due to residual spontaneous emission and collisions with background particles [22].

It is also possible to create a Schrödinger cat state of the target group that is not entangled to the control atom. This process can be considered as a transfer of the state of the control atom to the target group. The process works as follows. Initially, the control atoms are in state $(|\Uparrow\rangle+|\Downarrow\rangle)/\sqrt{2}$ and the target group in state $|\pi/2,0\rangle$. After the two-step protocol described above with $\mu=\pi/2$, the resulting state will be $(|\uparrow\rangle\otimes|\pi/2,\pi/2\rangle+|\downarrow\rangle\otimes|\pi/2,-\pi/2\rangle)/\sqrt{2}$. Next, a microwave $\pi/2$ pulse around $x$ axis is applied only to the control atom. The resulting state will be $[|\uparrow\rangle\otimes(|\pi/2,\pi/2\rangle-|\pi/2,-\pi/2\rangle)+|\downarrow\rangle\otimes(|\pi/2,\pi/2\rangle+|\pi/2,-\pi/2\rangle)]/\sqrt{2}$. Then the state of the original atom is detected. If the outcome is $|\uparrow\rangle$, then a $\pi/2$ pulse around $y$ axis is applied to the target group. If the outcome is $|\downarrow\rangle$, then a $-\pi/2$ pulse around $y$ axis is applied. In this way, the final sate of the target group will be $(|\Uparrow\rangle+|\Downarrow\rangle)/\sqrt{2}$, a Schrödinger cat state that is not entangled to the control atom.

## 4. Hidden order in a distorted squeezed state

It can be shown that $\mu=\pi/n$ ($n$ being an integer) is of special interest in elucidating the behavior OATS, especially under conditions when the Husimi QPD for the final state appears distorted. Since the behavior of OATS for $\mu=\pi/n$ is independent of the sign of $n$, we assume $n>0$ for convenience. We show in the Appendix that the OATS propagator with $\mu=\pi/n$ can be expressed as

$$e^{-i(\pi/n)S_z^2} = \frac{1}{\sqrt{n}} e^{-i\pi/4} \sum_{l=\alpha}^{n-1+\alpha} e^{i(\pi/n)l^2} e^{-i(2\pi/n)lS_z} \qquad (2)$$

which is a sum of $n$ rotational operators around the $z$-axis with their rotational angles forming an arithmetic progression, and $\alpha \equiv \{(N+n)/2\}$ ($\{\bullet\}$ means the fractional part) is 0 if $N+n$ is even and is $1/2$ if $N+n$ is odd. The well-known case of $n=2$ [25, 26, 27, 28] is just a special example of this property. The intermediate states between the OATS process and its inversion for $\mu = \pi/2$ and $\mu = \pi/3$ are plotted in Figure 3, using the Husimi QPDs. All possible final states along with the corresponding states of the control group are shown at the rightmost side. It should be noted that the state of the control group that gives a particular final state of the target group is not unique. The states of the control group shown in Figure 3 are what can be prepared with the fewest atoms.

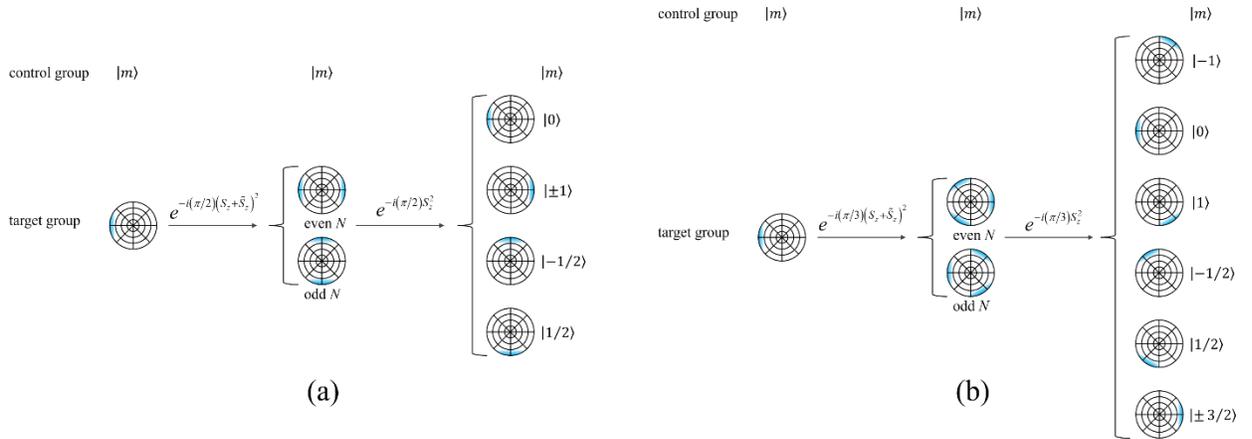

Figure 3 (a) Husimi QPDs of the states at each step of the protocol for $\mu = \pi/2$. The intermediate state of the target group after the OATS process is a superposition of two CSSs, separated by 180° azimuthally. The orientations of these two CSSs depend on the parity of $N$, which is the total number of atoms in both the control group and the target group. (b) Husimi QPDs of the states at each step of the protocol for $\mu = \pi/3$. The intermediate state of the target group after the OATS process is a superposition of three CSSs, separated by equal azimuthal angles. The orientations of the three CSSs depend on the parity of $N$.

A direct implication of this result is that after an OATS process with $\mu = \pi/n$, a CSS will evolve to a superposition of $n$ equally weighted CSSs, distributed with equal distances azimuthally. This may appear to contradict the fact that for $\mu \ll 1$, the Husimi quasi-probability (HQP) distribution of the state resulting from the OATS process appears as a single ellipse. However, there is no real contradiction, since for $\mu = (1/n) \ll 1$ the ellipse results from interference between the $n$ CSSs. In this regime, as $\mu$ increases from 0, the ellipse is first regular and then becomes distorted. It was believed that the states corresponding to distorted HQP distributions were chaotic. However, the result of Eq. (2) indicates that there are hidden orders in the states corresponding to distorted HQP distributions for $\mu = 1/n$. Figure 4 shows the HQP distributions of the quantum state with $N = 40$ produced by OATS for $\mu = 0$, $\pi/50$, $\pi/10$, 0.33 ($\approx \pi/9.5$), and $\pi/9$. The white circles show the approximate perimeters (defined in terms of the $1/e^2$ radii) of the HQP distributions of the CSS's constituting the quantum states. For small values of $\mu$, the HQP distribution is elliptical, like the $\mu = \pi/50$ case. As $\mu$ increases, the HQP distribution is stretched more and becomes distorted, like the cases of $\mu = \pi/10$, $0.33 (\approx \pi/9.5)$, and $\pi/9$. However, for the case of $\mu = \pi/n$ in this regime, the HQP distribution has clear nodes and antinodes, with the antinodes being in the HQP distributions of the CSS's constituting this state.

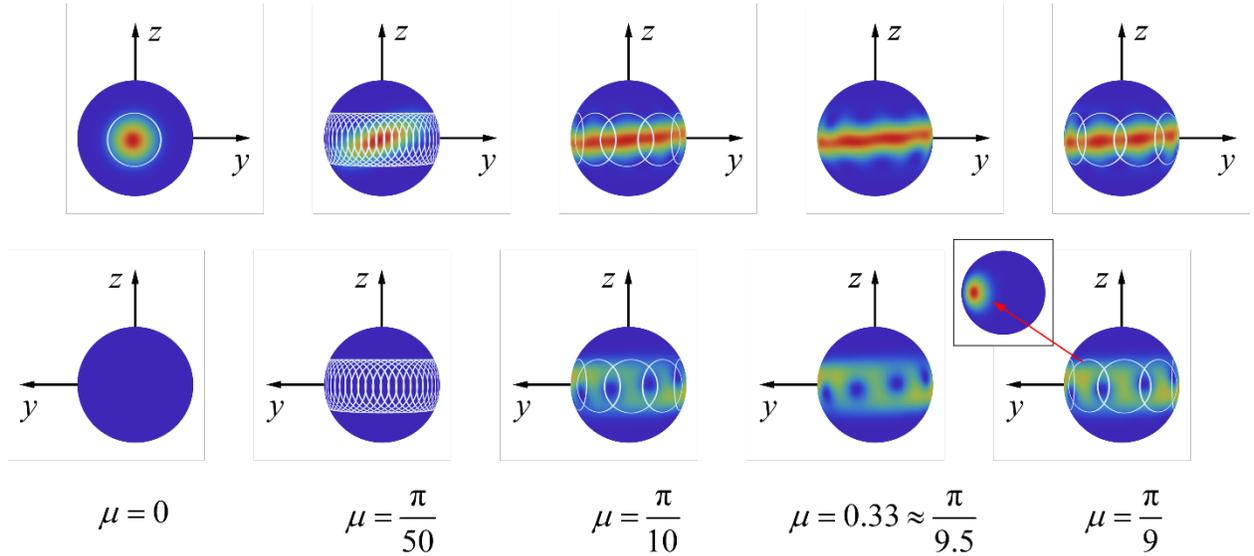

Figure 4 HQP distributions of the quantum state with $N = 40$ produced by OATS for $\mu = 0, \pi/50, \pi/10, 0.33 (\approx 0.95\pi)$, and $\pi/9$. The white circles show the approximate perimeters (defined in terms of the $1/e^2$ radii) of the HQP distributions of the CSS's constituting the quantum states.

## 5. Conclusion

To summarize, we have proposed a method of using OATS in an echo configuration to control a macroscopic scale ensemble of atoms with only one atom. With this method, we can produce a macroscopic superposition by entangling an ensemble to the control atom that is in a quantum superposition. This scheme mimics the true idea of the cat experiment proposed by Schrödinger and can be used to generate Schrödinger cat states deterministically and more reliably. In addition to the echo configuration, we have also investigated the behavior of one-axis-twist squeezing for some special values of squeezing parameter $\mu$. We find that the squeezing propagator can be expressed as the sum of $n$ rotation operators if $\mu = \pi/n$, where $n$ is a non-zero integer. A direct consequence of this property of one-axis-twist squeezing is that there is a hidden order in a squeezed state when $\mu = \pi/n$ even if its Husimi quasi-probability distribution looks distorted.

## Acknowledgement

This work has been supported equally in parts by the Department of Defense Center of Excellence in Advanced Quantum Sensing under Army Research Office grant number W911NF202076, and the U.S. Department of Energy, Office of Science, National Quantum Information Science Research Centers, Superconducting Quantum Materials and Systems Center (SQMS) under contract number DE-AC02-07CH11359.

# Appendix

We consider first the simplest nontrivial case $\mu = \pi/2$. An eigenstate of $S_z$ with the eigenvalue of $m$ is denoted as $|m\rangle$. It should be noted that $|m\rangle$ is degenerate because a non-degenerate state is defined by two quantum number: the total spin $S$ and the spin in the z-direction $m$. Because the OATS propagator does not contain the total spin operator $S$ but only the z-component of the spin operator $S_z$, the quantum number $S$ is irrelevant in this context. If the propagator $\exp(-i(\pi/2)S_z^2)$ acts on an arbitrary eigenstate of $S_z$, denoted as $|m\rangle$, the resulting state can be expressed as:

$$e^{-i(\pi/2)S_z^2}|m\rangle = e^{-i(\pi/2)m^2}|m\rangle \tag{3}$$

The simplified value of the expression $e^{-i(\pi/2)m^2}$ depends on whether $m$ is an integer or half-integer (i.e., whether $N$ is even or odd). If $m$ is an integer (i.e., $N$ is even), we then we have:

$$e^{-i(\pi/2)m^2} = (-i)^{m^2} = \begin{cases} -i, & m \text{ is odd} \\ 1, & m \text{ is even} \end{cases} = \frac{1}{\sqrt{2}}e^{-i\pi/4}\left[1+(-1)^m i\right] \tag{4}$$

Substituting this result into Eq. (3), we obtain the equation

$$e^{-i(\pi/2)S_z^2}|m\rangle = \frac{1}{\sqrt{2}}e^{-i\pi/4}\left[1+i(-1)^m\right]|m\rangle = \frac{1}{\sqrt{2}}e^{-i\pi/4}\left(1+ie^{-i\pi S_z}\right)|m\rangle \tag{5}$$

Noting that $|m\rangle$ is an arbitrary eigenstate of $S_z$ and any state can be decomposed into a superposition of eigenstates of $S_z$, we can see that the identity

$$e^{-i(\pi/2)S_z^2}|\psi\rangle = \frac{1}{\sqrt{2}}e^{-i\pi/4}\left(1+ie^{-i\pi S_z}\right)|\psi\rangle \tag{6}$$

holds for an arbitrary state $|\psi\rangle$. Therefore, it can be concluded that

$$e^{-i(\pi/2)S_z^2} = \frac{1}{\sqrt{2}} e^{-i\pi/4} \left(1 + ie^{-i\pi S_z}\right) \tag{7}$$

Consider next the case where $m$ is a half-integer (i.e., $N$ is odd). In this case, it is more convenient to calculate $\exp(i(\pi/2)S_z)\exp(-i(\pi/2)S_z^2)|m\rangle$ first, as follows:

$$e^{i(\pi/2)S_z} e^{-i(\pi/2)S_z^2} |m\rangle = e^{i\pi/8} e^{-i(\pi/2)(m-1/2)^2} |m\rangle \tag{8}$$

Noting that $(m-1/2)$ is an integer in this case, we can see that Eq. (8) has the same form as Eq. (3). Following the same steps from Eq. (3) to Eq. (7), we obtain

$$e^{-i(\pi/2)S_z^2} = \frac{1}{\sqrt{2}} e^{-i\pi/8} e^{-i(\pi/2)S_z} \left(1 - e^{-i\pi S_z}\right) \tag{9}$$

The state of the target group at each step of the protocol for $\mu = \pi/2$ in terms of the Husimi quasi-probability distribution is shown in Figure 3.

When $\mu = \pi/3$, the OATS propagator can be expressed as $U = \exp(-i(\pi/3)S_z^2)$. Then Eq. (3) is modified to be

$$e^{-i(\pi/3)S_z^2} |m\rangle = e^{-i(\pi/3)m^2} |m\rangle \tag{10}$$

The value of $\exp(-i(\pi/3)m^2)$ depends on whether $m$ is an integer or half-integer (i.e., whether $N$ is even or odd). If $m$ is a half-integer (i.e., $N$ is odd), we have:

$$e^{-i(\pi/3)m^2} = \begin{cases} e^{-i3\pi/4}, & \left(m+\dfrac{1}{2}\right)\bmod 3 = 2 \\ e^{-i\pi/12}, & \text{otherwise} \end{cases} \tag{11}$$

$$= \frac{1}{\sqrt{3}} e^{-i\pi/4} \left(1 + e^{i\pi/3} e^{-i(2\pi/3)m} + e^{i4\pi/3} e^{-i(4\pi/3)m}\right)$$

Following the same steps from Eq. (3) to Eq. (7), we obtain

$$e^{-i(\pi/3)S_z^2} = \frac{1}{\sqrt{3}} e^{-i\pi/4} \left(1 + e^{i\pi/3} e^{-i(2\pi/3)S_z} + e^{i4\pi/3} e^{-i(4\pi/3)S_z}\right) \tag{12}$$

If $m$ is an integer (i.e., $N$ is even), it is convenient to calculate $\exp(i(\pi/3)S_z)\exp(-i(\pi/3)S_z^2)|m\rangle$ first, as follows:

$$e^{i(\pi/3)S_z} e^{-i(\pi/3)S_z^2} |m\rangle = e^{i\pi/12} e^{-i(\pi/3)(m-1/2)^2} |m\rangle \tag{13}$$

Noting that $(m-1/2)$ is a half-integer, we can see that Eq. (13) has the same form as Eq. (10). Following the same steps from Eq. (10) to Eq. (12), we obtain

$$e^{-i(\pi/3)S_z^2} = \frac{1}{\sqrt{3}} e^{-i\pi/6} e^{-i(\pi/3)S_z} \left(1 + e^{i2\pi/3} e^{-i(2\pi/3)S_z} + e^{-i(4\pi/3)S_z}\right) \tag{14}$$

We consider next the general case $\mu = \pi/n$, where $n$ is a positive integer. We first consider the discrete function $\exp(i(\pi/n)k^2)$, where $k$ is the independent variable that is an integer if $n$ is even and a half-integer if $n$ is odd (i.e., the parity of $2k$ is the same as the parity of $n$). Particular examples of this periodicity can be seen from Eqs. (4) and (11). This function has a period of $n$ because

$$e^{i(\pi/n)(k+n)^2} = e^{i(\pi/n)(k^2+2kn+n^2)} = e^{i(\pi/n)k^2} e^{i2\pi n(k+n/2)} = e^{i(\pi/n)k^2} \tag{15}$$

The last step in this equation is based on the identity $\exp(i2\pi(k+n/2))=1$. Since the function $\exp(i(\pi/n)k^2)$ has a period of $n$, the summation $\sum_{k=k_0}^{k_0+(n-1)}\exp(i(\pi/n)k^2)$ has the same value for all values of $k_0$ that satisfy the condition that the parity of $2k_0$ is the same as the parity of $n$. The result of this summation is a type of quadratic Gauss sum [32], which equals $\sqrt{n}\exp(i\pi/4)$, i.e.,

$$\sum_{k=k_0}^{k_0+(n-1)} e^{i(\pi/n)k^2} = \sqrt{n}e^{i\pi/4} \tag{16}$$

for all the allowable values of $k_0$.

We next consider an arbitrary eigenstate of $S_z$, denoted as $|m\rangle$. The variable $k$ can be substitute with $(m-l)$, where $l$ is an integer (half-integer) variable if $(N+n)$ is even (odd) so that $(m-l)$ is an integer. Then Eq. (16) can be rewritten as

$$\sum_{l=\alpha}^{n-1+\alpha} e^{i(\pi/n)(m-l)^2} = \sqrt{n}e^{i\pi/4} \tag{17}$$

where $\alpha \equiv \{(N+n)/2\}$ ($\{\bullet\}$ means the fractional part) takes the value of 0 if $N+n$ is even and takes the value of $1/2$ if $N+n$ is odd. It follows from Eq. (17) that

$$\sqrt{n}e^{i\pi/4} = e^{i(\pi/n)m^2} \sum_{l=\alpha}^{n-1+\alpha} e^{i(\pi/n)l^2} e^{-i(2\pi/n)lm} \tag{18}$$

which gives

$$e^{-i(\pi/n)m^2} = \frac{1}{\sqrt{n}} e^{-i\pi/4} \sum_{l=\alpha}^{n-1+\alpha} e^{i(\pi/n)l^2} e^{-i(2\pi/n)lm} \tag{19}$$

Following the same steps from Eq. (3) to Eq. (7), we obtain

$$e^{-i(\pi/n)S_z^2} = \frac{1}{\sqrt{n}} e^{-i\pi/4} \sum_{l=\alpha}^{n-1+\alpha} e^{i(\pi/n)l^2} e^{-i(2\pi/n)lS_z} \quad (20)$$

which is the general result for arbitrary $N$ and $n$ shown in Eq. (2).

[1] Kovachy, T., Asenbaum, P., Overstreet, C., Donnelly, C. A., Dickerson, S. M., Sugarbaker, A., ... & Kasevich, M. A. (2015). Quantum superposition at the half-metre scale. Nature, 528(7583), 530-533.
[2] E. Schroedinger, "The present situation in quantum mechanics," Naturwissenschaften 23, 807 (1935); English translation by J. D. Trimmer, Proc. Am. Philos. Soc. 124(5), 323-338 (1980).
[3] Penrose, R. (1996). On gravity's role in quantum state reduction. General relativity and gravitation, 28(5), 581-600.
[4] Penrose, R. (2014). On the gravitization of quantum mechanics 1: Quantum state reduction. Foundations of Physics, 44(5), 557-575.
[5] Penrose, R. (2014). On the gravitization of quantum mechanics 2: Conformal cyclic cosmology. Foundations of Physics, 44(8), 873-890.
[6] Diósi, L. (1987). A universal master equation for the gravitational violation of quantum mechanics. Physics letters A, 120(8), 377-381.
[7] Diósi, L. (1989). Models for universal reduction of macroscopic quantum fluctuations. Physical Review A, 40(3), 1165.
[8] Bose, S., Jacobs, K., & Knight, P. L. (1997). Preparation of nonclassical states in cavities with a moving mirror. Physical Review A, 56(5), 4175.
[9] Marshall, W., Simon, C., Penrose, R., & Bouwmeester, D. (2003). Towards quantum superpositions of a mirror. Physical Review Letters, 91(13), 130401.
[10] Kleckner, D., Pikovski, I., Jeffrey, E., Ament, L., Eliel, E., Van Den Brink, J., & Bouwmeester, D. (2008). Creating and verifying a quantum superposition in a micro-optomechanical system. New Journal of Physics, 10(9), 095020.
[11] Fang, R., Sarkar, R., & Shahriar, S. M. (2020). Enhancing the sensitivity of an atom interferometer to the Heisenberg limit using increased quantum noise. JOSA B, 37(7), 1974-1986.
[12] Sarkar, R., Fang, R., & Shahriar, S. M. (2018). High-Compton-frequency, parity-independent, mesoscopic Schrödinger-cat-state atom interferometer with Heisenberg-limited sensitivity. Physical Review A, 98(1), 013636.
[13] Camara, A., Kaiser, R., & Labeyrie, G. (2014). Scaling behavior of a very large magneto-optical trap. Physical Review A, 90(6), 063404.
[14] Arecchi, F. T., Courtens, E., Gilmore, R., & Thomas, H. (1972). Atomic coherent states in quantum optics. Physical Review A, 6(6), 2211.
[15] Dicke, R. H. (1954). Coherence in spontaneous radiation processes. Physical review, 93(1), 99.
[16] Kitagawa, M., & Ueda, M. (1993). Squeezed spin states. Physical Review A, 47(6), 5138.
[17] Schleier-Smith, M. H., Leroux, I. D., & Vuletić, V. (2010). Squeezing the collective spin of a dilute atomic ensemble by cavity feedback. Physical Review A, 81(2), 021804.
[18] Leroux, I. D., Schleier-Smith, M. H., & Vuletić, V. (2010). Implementation of cavity squeezing of a collective atomic spin. Physical Review Letters, 104(7), 073602.
[19] Zhang, Y. L., Zou, C. L., Zou, X. B., Jiang, L., & Guo, G. C. (2015). Detuning-enhanced cavity spin squeezing. Physical Review A, 91(3), 033625.
[20] Sørensen, A. S., & Mølmer, K. (2002). Entangling atoms in bad cavities. Physical Review A, 66(2), 022314.
[21] Hosten, O., Engelsen, N. J., Krishnakumar, R., & Kasevich, M. A. (2016). Measurement noise 100 times lower than the quantum-projection limit using entangled atoms. Nature, 529(7587), 505-508.
[22] Li, J., da Silva, G. R., Kain, S., & Shahriar, S. M. (2022). A generalized echo squeezing protocol with near-Heisenberg limit sensitivity and strong robustness against excess noise and variation in squeezing parameter. arXiv preprint arXiv:2204.08681.
[23] Davis, E., Bentsen, G., & Schleier-Smith, M. (2016). Approaching the Heisenberg limit without single-particle detection. Physical review letters, 116(5), 053601.

[24] Hosten, O., Krishnakumar, R., Engelsen, N. J., & Kasevich, M. A. (2016). Quantum phase magnification. Science, 352(6293), 1552-1555.

[25] Mølmer, K., & Sørensen, A. (1999). Multiparticle entanglement of hot trapped ions. Physical Review Letters, 82(9), 1835.

[26] Leibfried, D., Barrett, M. D., Schaetz, T., Britton, J., Chiaverini, J., Itano, W. M., ... & Wineland, D. J. (2004). Toward Heisenberg-limited spectroscopy with multiparticle entangled states. Science, 304(5676), 1476-1478.

[27] Leibfried, D., Knill, E., Seidelin, S., Britton, J., Blakestad, R. B., Chiaverini, J., ... & Wineland, D. J. (2005). Creation of a six-atom 'Schrödinger cat' state. Nature, 438(7068), 639-642.

[28] Monz, T., Schindler, P., Barreiro, J. T., Chwalla, M., Nigg, D., Coish, W. A., ... & Blatt, R. (2011). 14-qubit entanglement: Creation and coherence. Physical Review Letters, 106(13), 130506.

[29] Giovannetti, V., Lloyd, S., & Maccone, L. (2004). Quantum-enhanced measurements: beating the standard quantum limit. Science, 306(5700), 1330-1336.

[30] Saffman, M., Walker, T. G., & Mølmer, K. (2010). Quantum information with Rydberg atoms. Reviews of modern physics, 82(3), 2313.

[31] Zeng, Y., Xu, P., He, X., Liu, Y., Liu, M., Wang, J., ... & Zhan, M. (2017). Entangling two individual atoms of different isotopes via Rydberg blockade. Physical Review Letters, 119(16), 160502.

[32] Murty, M., & Pathak, S. (2017). Evaluation of the quadratic Gauss sum. Evaluation, 86(1-2).
21